\documentclass[aps,prl,twocolumn,amsmath,amssymb]{revtex4}
\usepackage{dcolumn}
\usepackage{bm}
\usepackage{graphicx}
\begin{document}
\title{CHUF and `unfreezing' (or de-arrest) of kinetic arrest in magnetic shape memory alloys}
\author{P. Chaddah and A. Banerjee}
\affiliation{UGC-DAE Consortium for Scientific Research, University Campus, Khandwa Road, Indore-452001, Madhya Pradesh, India.}

\begin{abstract}
CHUF (cooling and heating in unequal field) protocol allows establishing phase coexistence through macroscopic measurements and enables distinguishing the metastable and stable phases (amongst the coexisting phase fractions across a first order magnetic transition) of a glass-like arrested state (GLAS). The observation of kinetic arrest in magnetic shape memory alloys has become a very active area of research. We highlight recent CHUF measurements on these materials that show behaviour similar to that reported earlier in manganites and other intermetallic alloys, emphasizing the generality. The martensitic transition provides an opportunity to study various properties as the CHUF protocol enables tuning of phase fractions, and controlled devitrification, of GLAS.    
\end{abstract}

\maketitle
We have noted recently that the kinetics of a first order transition is dictated by the time required for the latent heat to be extracted and, in addition to the well-known quenched metallic glasses, the kinetics of any first order transition (including long-range-order to long-range-order transitions) could thus be arrested \cite{comment}. When a first order magnetic transition is kinetically arrested, this glass-like arrested state (GLAS) has been referred to as a `magnetic glass' \cite{chatto, chaddah1, kranti,ban, roy, chaddah2, ban1, pallavi, sathe, lakh}. 

Phase-coexistence has been observed to persist in many half-doped manganites even at the lowest temperature. It has been shown that cooling in different values of magnetic field may aid or prevent this kinetic arrest, and the coexisting phase fraction at low temperature, at any measurement field H, can be tuned by cooling in an appropriate field and then changing the field isothermally to H \cite{ban}. Cooling in some field may allow the first order transition to be completed and the equilibrium state to be established, while cooling in a very different field may totally inhibit (or arrest) the transition. Since the first order transition occurs over length scales of the correlation length, it was argued \cite{chaddah1, kranti} that the broadened transition will be completed only partially for cooling field lying between two values (H$_1$ and H$_2$) of magnetic field. If the cooling field is below H$_1$ the transition will be completed (totally arrested) if the high-temperature phase was ferromagnetic (antiferromagnetic), and if the cooling field is above H$_2$ the transition will be totally arrested (completed) if the high-temperature phase was ferromagnetic (antiferromagnetic). It was argued \cite{chaddah1, kranti, ban} that by using unequal and appropriately chosen cooling and warming fields (H$_C$ and H$_W$), de-arrest (or devitrification) of the kinetically arrested GLAS could be caused, and this de-arrest would be seen for only one sign of (H$_C$ - H$_W$). Further heating would cause this de-arrested state to undergo the reverse magnetic transition, and this cooling and heating in unequal field (CHUF) protocol would show a reentrant transition. These phenomenological predictions, where the reentrant transition is seen only for positive (H$_C$ - H$_W$) when the high temperature phase was ferromagnetic, apply for a generic `magnetic glass'. By similar arguments, the GLAS would show a reentrant transition only for negative (H$_C$ - H$_W$) when the high temperature phase was antiferromagnetic. 

Kainuma's group has reported \cite{umetsu, ito, ito2, lee, umetsu2} measurements on Ni-Mn-In based alloys on cooling in various high magnetic fields, and warming in lower fields. In these studies, the low-temperature martensite phase has lower magnetization, and the preceding statements for a high-temperature ferromagnetic phase apply. They observe the ``unfreezing of the P+M coexisting state" \cite{umetsu} when warming in a small field of 0.05 Tesla. This devitrification observed in manganites with a charge-ordered ground state \cite{ban, chaddah2} has now been reproduced in detail, following the CHUF protocol, in Ni-Mn-Sn based alloys for a 4 Tesla warming field \cite{ban2, lakh2}. For cooling fields greater than 4 Tesla, devitrification is observed during warming, showing an austenite to martensite to austenite reentrant transition. The CHUF protocol also provided the added confirmation that when the cooling field is smaller than 4 Tesla the devitrification is not observed. Further, with a fixed cooling field of 3 Tesla when the austenite to martensite transition is completed only partly during cooling, warming in fields lower than 3 Tesla shows devitrification. On the other hand, warming in fields higher than 3 Tesla does not show devitrification. This has been observed for both Mi-Mn-In based alloys \cite{lakh3} and for Ni-Mn-Sn based alloys \cite{lakh4}. The analogy between the magnetic shape memory alloys currently being studied, and the manganites studied in great detail earlier, is thus complete. The magnetic shape memory alloys appear to comply with the predictions of the magnetic glass concept. The magnetic glass concept has been discussed for magnetic shape memory alloys by Sharma et al. \cite{sharma} also. But it has not been discussed in other extensive work on kinetic arrest of magnetic shape memory alloys \cite{umetsu, ito, ito2, lee, umetsu2, kara,ito3, chatt}. The CHUF protocol has not been used in magnetic shape memory alloys, except in the very recent works \cite{ban2, lakh2, lakh3, lakh4}.  

The applicability of CHUF to establish the tuneability of coexisting phase fractions with cooling field, and to establish the existence of a magnetic glass state has recently been also claimed in another family of materials showing kinetic arrest, viz. doped cobaltites \cite{rav}. It is also being applied to more materials in the manganite family \cite{rao, laks, doshi}. The martensitic transition, however, offers interesting observations through structural studies of, for example, nucleation and growth during devitrification \cite{ban3}. The presence of Sn in some of these magnetic shape memory alloys also affords the possibility of M$\ddot{o}$ssbauer measurements of this devitrification process under the CHUF protocol. There is a need to study the magnetic shape memory alloys in detail, as they may provide information on GLAS formed by arrest of first order magnetic transitions between structurally ordered phases.

\end{document}